# Spin-Orbit Torques in Transition Metal Dichalcogenide/Ferromagnet Heterostructures

**Jan Hidding[1*], Marcos H. D. Guimarães[1*]**

[1]Zernike Institute for Advanced Materials, University of Groningen, The Netherlands

**\* Correspondence:**
Corresponding Authors
m.h.guimaraes@rug.nl, jan.hidding@rug.nl

**Abstract**

In recent years, there has been a growing interest in spin-orbit torques (SOTs) for manipulating the magnetization in nonvolatile magnetic memory devices. SOTs rely on the spin-orbit coupling of a nonmagnetic material coupled to a ferromagnetic layer to convert an applied charge current into a torque on the magnetization of the ferromagnet (FM). Transition metal dichalcogenides (TMDs) are promising candidates for generating these torques with both high charge-to-spin conversion ratios, and symmetries and directions which are efficient for magnetization manipulation. Moreover, TMDs offer a wide range of attractive properties, such as large spin-orbit coupling, high crystalline quality and diverse crystalline symmetries. Although numerous studies were published on SOTs using TMD/FM heterostructures, we lack clear understanding of the observed SOT symmetries, directions, and strengths. In order to shine some light on the differences and similarities among the works in literature, in this mini-review we compare the results for various TMD/FM devices, highlighting the experimental techniques used to fabricate the devices and to quantify the SOTs, discussing their potential effect on the interface quality and resulting SOTs. This enables us to both identify the impact of particular fabrication steps on the observed SOT symmetries and directions, and give suggestions for their underlying microscopic mechanisms. Furthermore, we highlight recent progress of the theoretical work on SOTs using TMD heterostructures and propose future research directions.

## 1  Introduction

Spin-orbit torques (SOTs) are promising candidates for effective manipulation of magnetization through electric currents with applications in nonvolatile magnetic memory and logic devices. SOTs convert an electric current into a magnetic torque in non-magnetic/ferromagnetic heterostructure, i.e. an electric current through the stack can modulate the direction of the ferromagnet's magnetization [1] [2]. Devices showing large SOT efficiencies usually rely on a nonmagnetic material with large spin-orbit coupling in contact with a ferromagnet (FM). Transition metal dichalcogenides (TMDs), with chemical formula $MX_2$, where M is a transition metal (e.g. Mo, and W) and X a chalcogen element (e.g. S and Se), can provide large spin-orbit coupling and pristine surfaces which can result in a more intimate contact between the TMD and the FM layer. Furthermore, this family of materials offers a wide range of electronic and crystalline properties and symmetries. Although numerous articles were published on SOTs in TMD/ferromagnetic heterostructures, a clear understanding of the different mechanisms underlying observed SOTs remain yet to be understood.

In this mini-review, we give an overview of the recent progress on SOTs in TMD/FM heterostructures. Materials with high charge-to-spin conversion efficiencies, such as $WTe_2$ and $TaTe_2$ [3] [4] [5], are often considered as good candidates for large SOT efficiencies. However, large charge-to-spin conversion efficiencies are no guarantee for large SOT efficiencies, as SOTs are often an emergent phenomenon, depending on proximity effects (spin-orbit coupling and magnetic exchange), wavefunction overlap, and interface spin transparency (spin mixing conductance) as well. Indeed, the observed torques in TMD/FM heterostructures cannot always be explained by well-known effects such as the bulk spin Hall effect (SHE) [6] [7] [8] or the interfacial Rashba-Edelstein Effect (REE) [9] [10] [11] [12] [13] (Figure 1), indicating that other mechanisms involving material specific properties or interfacial effects are into play. This is supported by recent works suggesting that both the type of ferromagnetic layer [14] [15] and the interface properties between the TMD and the ferromagnetic layer [16] [17] [18] [19] [20] are of paramount importance for the observed SOTs, allowing for enhanced and unconventional SOTs.

To describe to different torques, we use the notation in terms of odd $\left(\tau_o^\zeta \propto \widehat{m} \times \hat{\zeta}\right)$ or even $\left(\tau_e^\zeta \propto \widehat{m} \times (\hat{\zeta} \times \widehat{m})\right)$ with respect to the magnetization direction ($\widehat{m}$), with $\zeta = x, y, z$. These torques are also named, respectively, field-like (FL) and damping-like (DL) torques in many papers in literature [2], with directions out-of-plane or in-plane with respect to the TMD/FM plane (Figure 1). For a fair comparison between the results in literature we use the torque conductivities $\left(\sigma_{o(e)}^\zeta\right)$ to quantify the SOT strength, which expresses the torques per unit area per unit electric field. This figure of merit is adopted rather than the torque efficiency ($\xi_{FL(AD)}^{j_c}$), because the electric field across the device can be more accurately determined when compared to the current density [21].

## 2  Discussion on recent progress

The field of SOTs using TMD-based devices has been rapidly developed in the past 5 years. Experimental studies have used different TMD sources (e.g. mechanical exfoliation or chemical vapor deposition, CVD), FM materials, deposition methods (e.g. sputtering or electron-beam evaporation), and measurement techniques, namely second-harmonic Hall (SHH) [22] [23] [24] [25] or spin-torque ferromagnetic resonance (ST-FMR) [26] [27] [28]. So far, it is unclear how these different techniques and procedures affect the measured SOTs.

In this section, we discuss the results for semiconducting, semi-metallic and metallic TMDs, giving an overview of their fabrication and measurement techniques (Table 1). Comparing the TMDs in this way allows us to pinpoint important differences and similarities in the observed torques.

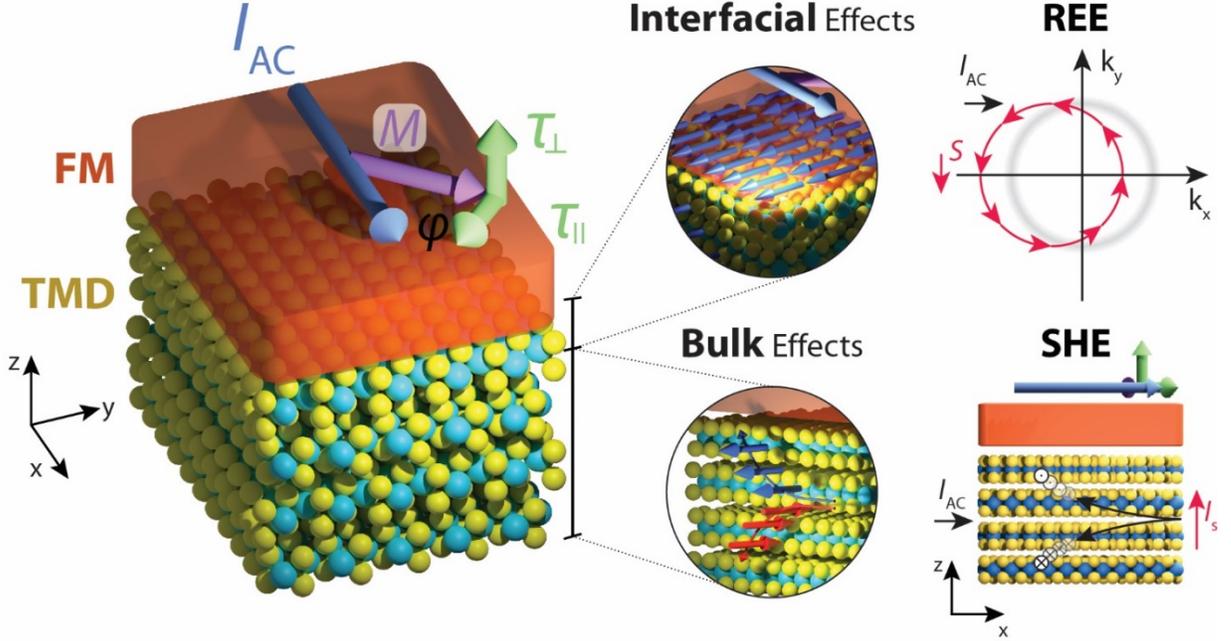

**Figure 1: Schematics of SOTs in TMD/FM heterostructures.** A charge current, usually oscillating at low (RF) frequencies for SHH (ST-FMR) measurements is applied along a device consisting of a TMD layer and a FM. The magnetization of the FM layer, oriented along an external magnetic field, observes a current-induced SOT in-plane ($\tau_\parallel$) and out-of-plane ($\tau_\perp$), indicated by the green arrows. These torques may arise from multiple microscopic effects arising in the bilayer, which may originate either from the TMD/FM interface (top), e.g. through the Rashba-Edelstein Effect (REE), or the bulk of the material (bottom), as for example through a spin Hall effect (SHE) in the TMD layer.

## 2.1 Semi-conducting TMDs

Shao *et al.* were one of the first to examine SOTs in TMD/FM heterostructures [29]. There, SOTs were quantified by the non-resonant SHH measurements on monolayer (1L) $MoS_2$ and $WSe_2$ coupled with CoFeB (3 nm). They observed a temperature independent out-of-plane FL torque $\tau_o^y$ ($\hat{m} \times \hat{y}$) for both devices with a corresponding torque conductivity of $\sigma_o^y = 2.88 \times 10^3 \, (\hbar/2e) \, (\Omega m)^{-1}$ and $5.52 \times 10^3 \, (\hbar/2e) \, (\Omega m)^{-1}$ for $MoS_2$ and $WSe_2$, respectively. No in-plane DL torque of the form $\tau_e^y \left( \hat{m} \times (\hat{y} \times \hat{m}) \right)$ was observed in either of their devices. This DL torque is observed in SOT measurements on Pt/Py bilayers and is often ascribed to the SHE [30]. Since the monolayer TMDs are much less conductive than the FM layer, the SOTs here are interfacial in nature, and the results point to the REE mechanism [31] [32] [33] [34].

Interestingly, in a concurrent work, Zhang *et al.* obtained different results using a high-frequency technique, ST-FMR, on 1L-$MoS_2$/Permalloy ($Ni_{80}Fe_{20}$ – Py) 5 nm [35]. There, they identified an in-plane DL $\tau_e^y$ ($\hat{m} \times (\hat{y} \times \hat{m})$) and an out-of-plane FL torque $\tau_o^y$ ($\hat{m} \times \hat{y}$). A torque ratio, $\tau_o^y/\tau_e^y = 0.19 \pm 0.01$ was obtained, indicating that $\tau_e^y$ dominates over $\tau_o^y$, in contrast to the results by Shao and co-workers. This result was repeated using different deposition techniques of the FM layer (sputtering or electron-beam deposition), indicating that the observed torque is independent on the Py deposition technique. The different measurement techniques used by the two groups could explain the discrepancy in the observed torques. However, it has been shown that the SOTs quantified by ST-FMR and SHH techniques agree within the experimental accuracy for several systems [36] [37] [38] [39].

The discrepancy between results for $MoS_2$/FM bilayers suggests that not only the spin-orbit material but also the type of ferromagnetic material (CoFeB *vs* Py) can play a significant role in the observed torques. This is theoretically substantiated in a recent work [14], where calculations on $MoSe_2$/Co, $WSe_2$/Co and $TaSe_2$/Co heterostructures were performed. They find that the hybridization of the Co wavefunctions with those of the TMDs leads to dramatic transmutation of the electronic and spin structure of the Co layers, even within eight layers away from the interface. This suggests that injecting unpolarized spin currents in these spin-orbit-proximitized layers of Co generates nonequilibrium spin densities, which in turn leads to a nonzero local torque on the magnetization. Both the spin polarization direction and magnitude were shown to differ between the different TMDs and complex spin textures were obtained for the spin-orbit-proximitized layers. These results indicate that the FM material can play an active role in the type of SOTs observed. Moreover, recent theoretical works [17] pointed out that different scattering mechanisms lead to different torque symmetries, indicating that the sample quality, symmetry and nature of scatterers also plays a role here. Different FM materials in FM/TMD heterostructures might therefore exhibit different SOTs as was the case for Shao *et al.* and Zhang *et al.*

More recently, $WS_2$ was studied by Lv *et al.* in a 1L-$WS_2$/Py (10 nm) heterostructure [40] using CVD-grown $WS_2$ and electron-beam evaporated Py layer. The authors observe both a DL torque $\tau_e^y$ and a FL torque $\tau_o^y$ in their ST-FMR measurements, which are ascribed to the interfacial REE. Furthermore, they observed a gate-dependent SOT ratio ranging from $\tau_o^y/\tau_e^y = 0.05$ to $0.22$ within a range of $V_g = -60$ V to $60$ V, absent in their reference sample of Pt/Py. Gate-dependent SOTs were observed in SHH measurements on a topological insulator [41], but not yet reported in TMD/FM heterostructures. The increasing SOT ratio with gate-voltage could be explained by an increased carrier density leading to an enhanced current at the $WS_2$/Py interface. The modulation of SOT strength using a gate voltage is a step towards applications for data storage and processing and more research should be done to improve the gate tunability of SOTs in TMD/FM heterostructures [42] [43] [44].

## 2.2 Semi-metallic TMDs

In addition to semiconducting TMDs, a variety of semi-metallic TMDs have been studied, with special focus given to low-symmetry crystals. A particularly interesting candidate is $WTe_2$, belonging to space group $Pmn2_1$. In a $WTe_2$/FM heterostructure, however, the symmetries are reduced to a single mirror plane perpendicular to the a-axis and the identity, space group $Pm$. The low device symmetry allows for unconventional SOTs, such as an out-of-plane DL torque $\tau_e^z$ ($\hat{m} \times (\hat{z} \times \hat{m})$), which is especially interesting for applications in high-density memory devices since these torques are very effective for magnetization switching of perpendicular magnetic anisotropy materials [45].

MacNeill *et al.* were the first to examine SOTs using $WTe_2$ [36]. Using ST-FMR, the authors observed $\tau_e^z$, along the conventional SOTs $\tau_o^y$ and $\tau_e^y$, and extracted a torque conductivity of $\sigma_e^z = 3.6 \pm 0.8 \times 10^3$ $(\hbar/2e)(\Omega m)^{-1}$ with the current driven along the low-symmetry a-axis. The other FL and DL torque conductivities were measured at $\sigma_o^y = 9 \pm 3 \times 10^3$ $(\hbar/2e)(\Omega m)^{-1}$ and $\sigma_e^y = 8 \pm$ $2 \times 10^3$ $(\hbar/2e)(\Omega m)^{-1}$, respectively. The magnitude of $\tau_e^z$ was found to depend on the angle between the electric current and the WTe$_2$ a-axis, showing a gradual decrease of the torque ratio $\tau_e^z/\tau_o^y$ when the projection of the current on the b-axis increases, giving support to its origin being correlated with the crystal symmetry. Even though an initial thickness dependence on the torques revealed little variation, a more thorough study with a wider thickness range ($t = 0.7 - 16$ nm) revealed additional bulk contributions to the SOTs in addition to the interfacial ones [37]. The thickness dependence of $|\tau_o^y|$, shows a strong increase with increasing WTe$_2$ thickness, suggesting it originates from an Oersted field produced by the current in the WTe$_2$ layer. The unusual out-of-plane DL torque $\tau_e^z$ shows a slowly decreasing magnitude with increasing thickness ($t \geq 4$ nm), while thinner layers show significant device-to-device variations. In the same work, the authors indicated that the in-plane DL torque $\tau_e^y$ possesses a similar thickness dependence as $\tau_e^z$. These torques remain large down a WTe$_2$ monolayer, suggesting that their microscopic origin is interfacial with some possible (smaller) additional bulk contribution.

Subsequent studies indicated a strong temperature dependence ($2 - 300$ K) on $\tau_o^y$ with the current flowing along the b-axis of WTe$_2$ using ST-FMR measurements [46]. While this temperature dependence was observed for thicker samples (20 nm and 31 nm), thinner samples (5.6 nm and 7.0 nm) only showed a weak temperature dependence. Furthermore, for a current applied along the a-axis (I//a), no temperature dependence is observed. A torque conductivity as high as $\sigma_o^y = 1.3 \times 10^5$ $(\hbar/2e)(\Omega m)^{-1}$ was reported. Calculations of the Oersted field contribution to $\tau_o^y$ could not explain the large enhancement. The enhanced SOT at low temperatures with I//b-axis was therefore ascribed to a spin accumulation created by spin-momentum locking in Fermi arcs which exist only along the b-axis, experimentally observed for WTe$_2$ nanoribbons with thicknesses in the range of 10 nm to 40 nm [47]. The origin of the relatively high $\tau_o^y$ which remains for thinner devices, is ascribed to the REE.

More recently, WTe$_2$/Py heterostructures have been shown to be very efficient for current-induced in-plane magnetization switching, with switching current densities in the order of $10^5$ A/cm$^2$ [39]. In the same work, the authors also reported a thickness dependence on the spin Hall efficiency in WTe$_2$, with larger values at higher thicknesses. However, the ST-FMR results show a significant frequency dependence and the role of artifacts such as skin-depth effects could not be ruled out. Nevertheless, the low threshold for current-induced magnetization switching indicates a promising direction for TMDs in future applications. Interestingly, these structures have also shown the presence of a Dzyaloshinskii-Moriya interaction, an essential ingredient for chiral magnetism.

The anisotropic in-plane conductivity in low-symmetry crystals can also impact SOTs. Results on TaTe$_2$/Py heterostructures have shown SOTs with Dresselhaus-like symmetries ($\hat{m} \times \hat{x}$) [38]. These torques have been shown to arise from Oersted-fields, generated by in-plane transverse current components due to conductivity anisotropy of TaTe$_2$. A similar, albeit smaller effect has been shown to be present in WTe$_2$/Py bilayers. Apart from the regular Oersted torque and Dresselhaus-like torque in the TaTe$_2$/Py heterostructures, the other torques are small or zero. Cross-sectional high-angle annular dark-field scanning transmission electron microscopy (HAADF-STEM) has indicated intermixing at the TaTe$_2$/Py interface which is likely to affect the effective SOTs due to a change in the local electronic environment and the spin mixing conductance of the interface. Interestingly, a change in the SOTs in topological-insulator/ferromagnet devices due to intermixing at the interface has been recently reported [48]. Here we point out that in addition to the changes in the SOTs arising from the different electronic structures for devices using different FM layers (e.g. Py, Co, CoFeB), the materials intermixing should also be carefully considered and potentially quantified in order to obtain a more in-depth understanding of the microscopic mechanisms involved.

Interestingly, both TaTe$_2$ and WTe$_2$ have shown to induce an in-plane magnetic anisotropy on Py, indicating a strong interaction between the semi-metallic TMDs and the FM layer. The anisotropy induced by WTe$_2$ was shown to be about 10's of mT and one order of magnitude larger than the one induced by TaTe$_2$. Additionally, the two TMDs induced anisotropy in different directions with respect to their crystal orientations, hinting towards the dependence of the induced magnetic anisotropy and the electronic structure of the TMD.

Another interesting semi-metallic TMD is $\beta$-MoTe$_2$ which, different than WTe$_2$ and similar to TaTe$_2$, possess inversion symmetry in its bulk form. Using $\beta$-MoTe$_2$/Py bilayers Stiehl *et al.* observe the presence of an out-of-plane DL torque $\tau_e^z$ using ST-FMR measurements [49]. This is allowed by the inversion symmetry breaking at the $\beta$-MoTe$_2$/Py interface and indicates that inversion asymmetry in the bulk is not a strict requirement for $\tau_e^z$ to be observed. The authors report a thickness independent torque conductivity of $\sigma_e^z = 1.02 \pm 0.03 \times 10^3$ $(\hbar/2e)(\Omega m)^{-1}$, 1/3 of the value reported for WTe$_2$. The standard in-plane DL torque $\tau_e^y$ was also observed with $\sigma_e^y = 5.8 \pm 0.16 \times 10^3$ $(\hbar/2e)(\Omega m)^{-1}$, and showed no apparent thickness dependence. The lack of a thickness dependent on $\tau_e^z$ and $\tau_e^y$ for both WTe$_2$ and $\beta$-MoTe$_2$, strongly suggests an interfacial origin for these SOTs.

In addition to the out-of-plane DL torque $\tau_e^z$, the low crystal symmetries of WTe$_2$ and $\beta$-MoTe$_2$ also allow for the presence of an in-plane FL torque $\tau_o^z$ ($\hat{m} \times \hat{z}$). While this torque was not observed in WTe$_2$, it was present in $\beta$-MoTe$_2$ devices. There, both $\tau_e^z$ and $\tau_o^z$ have shown similar temperature dependences, but different thickness dependences, hinting towards two microscopic mechanisms for $\tau_o^z$: one related and another unrelated to $\tau_e^z$. However, the physical mechanisms that generate these torques are still unknown.

More recently, PtTe$_2$/Py devices [50] have shown a high spin-torque conductivity for the in-plane DL torque $\sigma_e^y = 1.6 \times 10^5$ $(\hbar/2e)(\Omega m)^{-1}$. This value is one order of magnitude (or larger) than the values encountered in other TMD-based devices and comparable to devices based on heavy-metal or topological-insulators. This large spin-torque conductivity has been ascribed to a combination of the SHE and spin-momentum locking in topological surface states of PtTe$_2$, as previously observed in topological insulators [51] [52] [53] [54].

## 2.3 Metallic TMDs

Despite offering stronger spin-orbit interaction and higher conductivity, metallic TMDs have received less attention than their semi-metallic and semiconducting counterparts. To date, only two experimental studies have been reported [55] [56].

Thickness dependent ST-FMR measurements on NbSe$_2$ (1 to 10 layers) /Py heterostructures revealed an in-plane DL torque $\tau_e^y$ with a torque conductivity ($\sigma_e^y = 3 \times 10^3$ $(\hbar/2e)(\Omega m)^{-1}$) comparable to other TMD/Py heterostructures and observable down to a monolayer of NbSe$_2$ [55]. Similar to $\beta$-MoTe$_2$/Py [49], $\tau_e^y$ shows only a weak thickness dependence. An out-of-plane FL torque ($\sigma_o^y = 40 \times 10^3$ $(\hbar/2e)(\Omega m)^{-1}$) was also observed, and attributed to Oersted fields due to their linear scaling with NbSe$_2$ thickness. However, for thin NbSe$_2$ layers, the estimated Oersted-field contribution overestimates the observed torque magnitude, and, for monolayer NbSe$_2$ a sign change is observed. These observations for $\tau_e^y$ and $\tau_o^y$ indicate a contribution from interfacial torques.

In addition to the SOTs with conventional symmetries, an in-plane FL torque $\tau_o^z$ ($\hat{m} \times \hat{z}$) was observed in some devices. Since the trigonal symmetry of NbSe$_2$ does not allow for their presence, and given the seemly random thickness dependence of $\tau_o^z$, the authors argue that these torques could arise from uncontrollable strain from the fabrication procedure, which reduces the NbSe$_2$ symmetries. Although $\tau_e^z$ is subject to the same symmetry constraints, $\tau_e^z = 0$ for all measured devices, which is in contrast to the torques obtained for WTe$_2$, where $\tau_o^z = 0$, and $\tau_e^z \neq 0$ [36] [37]. This indicates that symmetry analysis alone is not sufficient to predict the observed torques in these systems and that other microscopic factors related to, for example, interface quality [23] [17], Berry curvature [57], or local atomic point-group symmetries [58] could play an important role.

A large spin-torque conductivity of $\sigma_e^y = 2.63 \times 10^5 (\hbar/2e)(\Omega m)^{-1}$ has been recently reported for the metallic monolayer TaS$_2$/Py heterostructures [56] using ST-FMR measurements. This result is attributed to a clean interface which is supported by cross-sectional TEM imaging. Using DFT calculations, the authors observe a considerable redistribution of the band structure which they hold accountable for the prominent DL torque.

## 3    Conclusions

In this review, we have given an overview of the current status of the field of SOTs in TMD/FM heterostructures. A multitude of SOT symmetries, magnitudes and directions were observed, which could not always be explained by well-known effects such as the SHE and REE. Different mechanisms that do not rely on a large spin-orbit coupling, such as anisotropic in-plane conductivity and uniaxial strain, can also play an important role. Additionally, interfacial effects such as spin-orbit filtering, spin-orbit precession and spin-momentum locking in topological surface states may affect the observed torques. In combination with the large torque conductivities obtained at clean interfaces, this suggests that the TMD/FM interface quality is of paramount importance for both the torque magnitude and direction. Lastly, the ferromagnetic layer, often considered to play a passive role, can have a significant effect on the observed SOTs due to changes of the electronic structure and intermixing at the interface. Dzyaloshinskii–Moriya interaction (DMI) has been shown to arise at TMD/FM interfaces demonstrating a strong interaction between these materials [59] [39] [60]. The large interfacial DMI in these heterostructures could be explored in future devices combining chiral magnetic structures and SOTs.

Although the crystal symmetry allows for a reasonable prediction of the allowed SOTs, a better understanding of the underlying microscopic mechanisms is key in qualitatively explaining the observed SOTs. In this regard, thickness dependent measurements provide a tool to better differentiate bulk effects from interfacial effects. However, as the contributions of different effects are measured all at once, it remains difficult to distinguish the numerous mechanisms underlying the torques with the current experimental techniques. To clarify the role of the ferromagnetic layer, a variety of devices with different FM materials should be fabricated.

Van der Waals heterostructures composed of TMDs, two-dimensional magnetic materials and graphene should allow for the study of SOTs at the ultimate thickness. Due to their small thickness, in addition to possibly reducing the device footprint, atomically-thin materials are more susceptible to external stimuli, such as gate-voltages, strain and illumination. Along these lines, interesting predictions point to the modulation of SOT and magnetization by gate-voltages in these structures [61] [62]. The exploration of gate-tunable SOTs in TMD/FM heterostructures could serve as a first step towards non-volatile data processing and storage as well as processing-in-memory applications. By giving an overview of the current status of the field, we hope to facilitate progress on elucidating the different underlying physical mechanisms for the SOTs.

## 4      Acknowledgements

We acknowledge funding from the Dutch Research Council (NWO) Start-Up Grant (STU.019.014), the European Union Horizon 2020 research and innovation program under grant agreements No 696656 and 785219 (Graphene Flagship Core 2 and Core 3), and the Zernike Institute for Advanced Materials.

**Table 1: Recent studies on TMD/FM heterostructures with their fabrication techniques and spin torque conductivities.**

| Reference | SOT Material (thickness) | Fabrication Technique | Ferro-magnet | Deposition Technique | Measurement Technique | Spin torque conductivity $[\times 10^3 \, (\hbar/2e)(\Omega m)^{-1}]$ | Proposed Mechanism / Source |
|---|---|---|---|---|---|---|---|
| **Semiconducting** | | | | | | | |
| [29] | $MoS_2$ (1L) | CVD | CoFeB (3 nm) | Magnetron sputtering | SHH | $\sigma_o^y = 2.88$ | REE |
| [29] | $WeS_2$ (1L) | CVD | CoFeB (3 nm) | Magnetron sputtering | SHH | $\sigma_o^y = 5.52$ | REE |
| [35] | $MoS_2$ (1L) | CVD | Py (5 nm) | Magnetron sputtering | ST-FMR | $\sigma_e^y = $ Observed | Interfacial |
| [40] | $WS_2$ (1L) | CVD | Py (10 nm) | E-beam evaporation | ST-FMR | $\sigma_o^y = $ Observed<br>$\sigma_e^y = $ Observed | REE<br>REE |
| **Semi-metallic** | | | | | | | |
| [36] | $WTe_2$ (1.8 nm – 15 nm) | Exfoliation | Py (6 nm) | Sputtering | ST-FMR/SHH | $\sigma_o^y = 9 \pm 3$<br>$\sigma_e^y = 8 \pm 2$<br>$\sigma_e^z = 3.6 \pm 0.8$<br>$\sigma_o^z = 0$ | Interfacial<br>Interfacial<br>Interfacial<br>- |
| [37] | $WTe_2$ (1L – 16 nm) | Exfoliation | Py (6 nm) | Sputtering | ST-FMR/SHH | $\sigma_o^y = $ Observed<br>$\sigma_e^y = $ Observed<br>$\sigma_e^z = $ Observed<br>$\sigma_o^z = 0$ | Oersted<br>-<br>-<br>- |
| [46] | $WTe_2$ (5.6 – 31 nm) | Exfoliation | Py (6 nm) | Sputtering | SHH | $\sigma_o^y = 1.3 \times 10^2$<br>$\sigma_e^y = $ Observed | Fermi arcs<br>- |
| [39] | $WTe_2$ (5.8 nm – 122 nm) | Exfoliation | Py (6 nm) | Sputtering | ST-FMR/SHH | $\sigma_e^y = 6 \times 10^1$ (I//b)<br>$\sigma_e^y = 5.95$ (I//a)<br>$\sigma_e^z = $ Observed (I//a) | Bulk |
| [38] | $TaTe_2$ (4.5 nm – 19.7 nm) | Exfoliation | Py (6 nm) | Sputtering | ST-FMR/SHH | $\sigma_e^y = $ Weak<br>$\sigma_o^z = $ Sometimes observed<br>$\sigma_e^z = 0$<br>$\sigma_o^x = $ Observed (Dresselhaus) | -<br>-<br>-<br>Oersted (resist. anisotropy) |
| [49] | $MoTe_2$ (1L – 14.2 nm) | Exfoliation | Py (6 nm) | Sputtering | ST-FMR | $\sigma_o^y = 15$ (Oersted)<br>$\sigma_e^y = 5.8 \pm 0.16$<br>$\sigma_e^z = 1.02 \pm 0.03$<br>$\sigma_o^z = 0.81 \pm 0.05$ (t>3 nm) | Oersted<br>Interfacial<br>Interfacial<br>Interfacial |
| [50] | $PtTe_2$ (3 nm – 20 nm) | CVD | Py (2.5, 5.0, 7.5, 10 nm) | Sputtering | ST-FMR | $\sigma_o^y = $ Observed<br>$\sigma_e^y = 1.6 \times 10^2$ | -<br>SHE + TSS |
| **Metallic** | | | | | | | |
| [55] | $NbSe_2$ (1L-10L) | Exfoliation | Py (6 nm) | Sputtering | ST-FMR | $\sigma_o^y = 40$<br>$\sigma_e^y = 3$<br>$\sigma_e^z = 0$<br>$\sigma_o^z = 1$ | Oersted<br>REE<br>-<br>Strain |
| [56] | $1T-TaS_2$ (1L) | Ion-beam Sputtering | Py | - | ST-FMR/SHH | $\sigma_o^y = $ Negligible<br>$\sigma_e^y = 2.63 \times 10^2$ | -<br>Interfacial |